\documentclass[preprint,aps,floats,nofootinbib]{revtex4}

\input epsf.tex
\def\DESepsf(#1 width #2){\epsfxsize=#2 \epsfbox{#1}}

\newcommand{\be}{\begin{equation}}
\newcommand{\ee}{\end{equation}}
\newcommand{\bea}{\begin{eqnarray}}
\newcommand{\eea}{\end{eqnarray}}

\def\thebibliography#1{\centerline{\bf REFERENCES}
  \list{[\arabic{enumi}]}{\settowidth\labelwidth{[#1]}\leftmargin
  \labelwidth\advance\leftmargin\labelsep\usecounter{enumi}}
\def\newblock{\hskip .11em plus .33em minus -.07em}\sloppy
  \clubpenalty4000\widowpenalty4000\sfcode`\.=1000\relax}

\begin{document}
\draft

\vspace*{0.5cm}

\title{Strong phase shifts and color-suppressed tree amplitudes \\
in $B \to DK^{(*)}$ and $B \to D\pi$, $D\rho$ decays}

\author{ \vspace{0.5cm}
C.~S.~Kim$^{1,2}$\footnote{cskim@yonsei.ac.kr}, ~
Sechul~Oh$^{2,3}$\footnote{scoh@phya.yonsei.ac.kr}, ~
and ~
Chaehyun~Yu$^1$\footnote{chyu@cskim.yonsei.ac.kr}}

\affiliation{ \vspace{0.3cm}
$^1$Department of Physics and IPAP, Yonsei University,
Seoul 120-479, Korea \\
$^2$Physics Division, National Center for Theoretical Sciences,
Hsinchu 300, Taiwan \\
$^3$Natural Science Research Institute, Yonsei University,
Seoul 120-479, Korea
\vspace{1cm}}

\vspace*{0.5cm}

\begin{abstract}
We analyze the decay processes $B \to DK$, $DK^*$, $D\pi$, and $D\rho$
in a model-independent way.
Using the quark diagram approach, we determine the magnitudes
of the relevant amplitudes and the relative strong phase shifts.
In order to find the most likely values of the magnitudes and the relative
strong phases of the amplitudes in a statistically reliable way, we use
the $\chi^2$ minimization technique.
We find that the strong phase difference between the color-allowed and the
color-suppressed tree amplitude can be large and is non-zero at 1$\sigma$
level with the present data.
The color-suppressed tree contributions are found to be sizably enhanced.
We also examine the validity of factorization and estimate the breaking
effects of flavor SU(3) symmetry in $B \to DK$, $D\pi$ and in $B \to DK^*$,
$D\rho$.
\end{abstract}
\maketitle

\section{Introduction}

A tremendous amount of experimental data on $B$ meson decays are being
collected from $B$ factory experiments, such as Belle and BaBar.
Experimentally plenty of two-body hadronic $B$ decays have been observed
and a lot of theoretical works on these decay processes have been done.
In particular, the first observation of the color-suppressed decay processes
$\bar B^0 \to D^0 \bar K^0$ and $\bar B^0 \to D^0 \bar K^*$ by the Belle
Collaboration \cite{Krokovny:2002ua} has drawn special attentions, since it allows one
to do a complete isospin analysis of the $B \to D K^{(*)}$ modes together
with the previously observed charged modes of the $B \to D K^{(*)}$ type.

Two-body hadronic $B$ meson decays to $D K^{(*)}$ and $D \pi$ final states
have been of great interest.  In these decay modes, there is no contribution
from penguin diagrams so that theoretical uncertainties involved in the
relevant QCD dynamics become much less.
Thus, these modes serve as a good testing ground for various theoretical
issues in hadronic $B$ decays, such as factorization hypothesis and final-state
interactions.  These processes are also expected to be useful for a
determination of the CP violating phases, e.g., $\phi_3$
\cite{Atwood:1996ci,Gronau:1998un,Jang:1998iz,Kim:2000ev}.

It has been expected that in a heavy quark limit, certain two-body
charmed $B$ decays, such as $\bar B^0 \to D^{(*)} \pi^-$ [referred
to as the class-1 (color-allowed) topology], can be explained well
with the factorization hypothesis implying small final-state
interactions. It has been confirmed in the QCD factorization
approach \cite{Beneke:2000ry}. However, in a recent work based on
the perturbative QCD (PQCD) approach \cite{Keum:2003js}, it was pointed out that
in order to explain $\bar B^0 \to D^0 \pi^\circ$ [referred to as the
class-2 (color-suppressed) topology] as well as $B^- \to D^0
\pi^-$ [referred to as the class-3 (involving both color-allowed
and color-suppressed) topology], there must exist a sizable
relative strong phase $\delta_{12}$ between the class-1 and the
class-2 amplitudes: e.g., $\delta_{12} = 59^0$. This relative
strong phase arises from QCD dynamics through short-distance
strong interactions and differs from the final-state strong phases
through long-distance rescattering interactions.

Motivated by experimental measurements of the branching ratios (BRs) for
$B \to D \pi$ and $B \to D K$ decays, some phenomenological studies have been
performed to determine the possible final-state rescattering strong phases
in these processes \cite{Xing:2001nj,Neubert:2001sj,
Chiang:2002tv,Cheng:2001sc,Xing:2003fe}.
Especially, in Refs. \cite{Xing:2001nj,Xing:2003fe} the $B \to D \pi$
and $B \to D K^{(*)}$ modes were studied {\it through the isospin analysis}.
However, in those works, the possibility of a sizable relative strong
phase between the color-allowed and the color-suppressed tree amplitudes was
completely ignored.
On the other hand, in Ref. \cite{Chiang:2002tv}, the $B \to D \pi ~(D \rho)$ and
$B \to D K^{(*)}$ modes were analyzed {\it in the topological quark diagram
approach} and in that analysis {\it flavor SU(3) symmetry} was assumed to combine
the relevant amplitudes with each other.

In this work, we re-analyze the $B \to DK$ and $B \to D \pi$ modes
as well as $B \to DK^*$ and $B \to D \rho$ in the quark diagram
approach, focusing on the following interesting issues. (i) We
estimate, in a model-independent way, the magnitude of the
relative strong phases, {\it taking into account the possibility
of  a sizable relative strong phase between the color-allowed and
the color-suppressed tree amplitudes}. This approach is different
from that by Xing \cite{Xing:2001nj,Xing:2003fe}, where the strong
phase difference between the color-allowed and the
color-suppressed tree amplitude was assumed to be zero. (ii) We
first study the $B \to DK^{(*)}$ and $B \to D\pi ~(D\rho)$
independently,  without using the flavor SU(3) symmetry, in
order to avoid the possibly large effect of SU(3) breaking.  (In
fact, we shall see later that the SU(3) breaking effect can be
sizable.) (iii) To determine the most likely values of the
magnitudes of the relative strong phase shifts  in a
statistically reliable way, we do the $\chi^2$ analysis (with the
flavor SU(3) and its breaking effect together) and explicitly show
that the relative final-state strong phases in $B \to DK$ and $B
\to D\pi$ are {\it non-zero at 1$\sigma$ level}. (iv) We examine
the validity of factorization approximation in these $heavy \to
heavy$ type decay modes, and estimate the flavor SU(3) symmetry
breaking effects in a model-independent way.

The paper is organized as follows.  The decay modes $B \to DK$ and $B \to DK^*$
are studied in Sec. II and the modes $B \to D\pi$ and $B \to D\rho$ are
analyzed in Sec. III.  In Sec. IV, the $\chi^2$ analysis using
$B \to DK$, $D\pi$ and $B \to DK^*$, $D\rho$ is presented.  The breaking effects
of the flavor SU(3) symmetry are estimated in Sec. V.
We conclude the analysis in Sec. VI.

\section{$B \to D K$ and $B \to D K^*$ decay modes}

First, let us consider the decay processes $B \to D K$.
The decay amplitudes for two-body hadronic $B$ decays can be represented
in terms of the basis of topological quark diagram contributions \cite{gronau},
such as $T$ (color-allowed tree amplitude), $C$ (color-suppressed tree amplitude),
$E$ (exchange amplitude), and so on.
The relevant decay amplitudes for $B \to DK$ can be written as
\begin{eqnarray}
A^{DK}_{0-} &\equiv& {\cal A}(B^- \to D^0 K^-)
 = T^{DK} + C^{DK} ,  \nonumber \\
A^{DK}_{+-} &\equiv& {\cal A}(\bar B^0 \to D^+ K^-)
 = T^{DK} ,  \nonumber \\
A^{DK}_{00} &\equiv& {\cal A}(\bar B^0 \to D^0 \bar K^0)
 = C^{DK} ,
\label{decayandtopo}
\end{eqnarray}
where the topological amplitudes $T^{DK}$ and $C^{DK}$ are defined as
\begin{eqnarray}
X^{DK} \equiv |X^{DK}| e^{i \delta_X}
 \equiv |V_{cb} V_{us}^*| a_X e^{i \delta_X},  ~~ (X = T,C)
\label{topoamp}
\end{eqnarray}
with the real amplitudes $a_{T (C)}$ and the strong phases
$\delta_{T(C)}$.
Note that no weak phase appears in the above amplitudes due to the
Cabibbo-Kobayashi-Maskawa (CKM) factor $V_{cb} V_{us}^*$.

\mbox{}From (\ref{decayandtopo}), the magnitudes $|T^{DK}|$ and
$|C^{DK}|$ and strong phase difference $(\delta_T -\delta_C)^{DK}$ of the
topological amplitudes can be determined in a model-independent way:
\begin{eqnarray}
|T^{DK}| &=& |A^{DK}_{+-}|
 = m_B \sqrt{{{8 \pi} \over {p_{_{DK}} \tau_0}} {\cal B}^{DK}_{+-}},  \nonumber \\
|C^{DK}| &=& |A^{DK}_{00}|
 = m_B \sqrt{{{8 \pi} \over {p_{_{DK}} \tau_0}} {\cal B}^{DK}_{00}},  \nonumber \\
\cos (\delta_T -\delta_C)^{DK}
 &=& { {|A^{DK}_{0-}|^2 - |A^{DK}_{+-}|^2 - |A^{DK}_{00}|^2}
  \over {2 |A^{DK}_{+-}| \cdot |A^{DK}_{00}|} }
 = { {(\tau_0 / \tau_-) {\cal B}^{DK}_{0-} -{\cal B}^{DK}_{+-} -{\cal B}^{DK}_{00}}
  \over {2 \sqrt{{\cal B}^{DK}_{+-} ~{\cal B}^{DK}_{00}}} },
\label{topo}
\end{eqnarray}
where $\tau_{-} ~(\tau_{0})$ is the life time of $B^- ~(\bar B^0)$.
The magnitude of the momentum $p_{_{DK}}$ of the $D(K)$ meson in the center
of mass frame is given by
\begin{eqnarray}
p_{_{DK}} = {1 \over 2 m_B} \sqrt{ [m_B^2 - (m_D +m_K)^2]
 [m_B^2 - (m_D -m_K)^2] }.
\end{eqnarray}
Notice that $(\delta_T -\delta_C)^{DK}$ is the relative strong phase of
the {\it color-suppressed} tree amplitude to the {\it color-allowed} tree
amplitude.

Since the same relations (\ref{decayandtopo}) also hold for the corresponding
$B \to D K^*$ modes, the above result in (\ref{topo}) can be used for
the relevant modes $B^- \to D^0 K^{*-}$, $\bar B^0 \to D^+ K^{*-}$ and
$\bar B^0 \to D^0 \bar K^{*0}$ by simply replacing $K$ by $K^*$.

\begin{table}
\caption{The BRs of $B\to DK$, $D K^*(892)$, $D \pi$, and $D \rho$
modes in units of $10^{-4}$. }
\smallskip
\begin{tabular}{|c|c||c|c|}
\hline
Mode & Experimental value & Mode & Experimental value  \\
\hline
~~$B^- \to D^0 K^- $~~ &~~ $3.7 \pm 0.6$
  ~~ &~~ $B^- \to D^0 K^{*-} $~~ &~~ $6.1 \pm 2.3$  \\
~~$\bar{B}^0 \to D^+ K^- $~~ &~~ $2.0 \pm 0.6$
  ~~ &~~ $\bar{B}^0 \to D^+ K^{*-} $~~ &~~ $3.7 \pm 1.8$  \\
~~$\bar{B}^0 \to D^0 \bar{K}^0 $~~ &~~ $0.50^{+0.13}_{-0.12}\pm 0.06$
  ~~ &~~ $\bar{B}^0 \to D^0 \bar{K}^{\ast 0}$
  ~~ &~~ $0.48^{+0.11}_{-0.10} \pm 0.05$ \\
\hline
~~$B^- \to D^0 \pi^- $~~ &~~ $49.7 \pm 3.8$
  ~~ &~~ $B^- \to D^0 \rho^- $~~ &~~ $134 \pm 18$  \\
~~$\bar{B}^0 \to D^+ \pi^- $~~ &~~ $26.8 \pm 2.9$
  ~~ &~~ $\bar{B}^0 \to D^+ \rho^- $~~ &~~ $78 \pm 14$ \\
~~$\bar{B}^0 \to D^0 \pi^0 $~~ &~~ $2.92 \pm 0.45$
  ~~ &~~ $\bar{B}^0 \to D^0 \rho^0 $~~ &~~ $2.9 \pm 1.0 \pm 0.4$ \\
\hline
\end{tabular}
\end{table}

\begin{table}
\caption{The numerical results for $|T|$, $|C|$, $|C/T|$, and
$\cos(\delta_T -\delta_C)$. The results shown in the last two columns
are obtained from the $\chi^2$ fit for $(\chi^2_{min} +1)$. }
\smallskip
\begin{tabular}{|c|c|c|c|c|}
\hline
Mode & $|T|$ ($10^{-7}$) & $|C|$ ($10^{-7}$) & $|C/T|$ & $\cos(\delta_T -\delta_C)$ \\
\hline
$B \to DK$ & $1.35 \sim 1.85$ & $0.68 \sim 0.92$ & $0.42 \sim 0.56$ & $0.03 \sim 0.73$ \\
$B \to D\pi$ & $4.4 \sim 6.8$ & $1.7 \sim 3.8$ & $0.51 \sim 0.69$ & $0.02 \sim 0.73$ \\
$B \to DK^*$ & $1.60 \sim 2.73$ & $0.68 \sim 0.90$ & $0.31 \sim 0.40$ & $0.1 \sim 1.0$ \\
$B \to D\rho$ & $7.9 \sim 11.6$ & $1.3 \sim 4.1$ & $0.32 \sim 0.41$ & $0.1 \sim 1.0$ \\
\hline
\end{tabular}
\end{table}

The experimental results on the BRs of $B \to DK$ and $DK^*$ as well as
$B \to D\pi$ and $D\rho$ are shown in Table I.
Using the measured BRs for $B \to DK$ decays, we calculate the magnitudes
of the color-allowed and the color-suppressed tree amplitudes and present
the results in Table II.
In Fig. 1, we show $\cos (\delta_T -\delta_C)^{DK}$ versus $|C^{DK}/T^{DK}|$.
Due to the large uncertainty in the present data, it is still
possible that the phase difference $(\delta_T -\delta_C)^{DK}$
vanishes.
But, for the central values of the experimental data,
\begin{eqnarray}
(\delta_T -\delta_C)^{DK} &=& 63.0^\circ ~, ~~{\rm or}~~ 297.0^\circ ~,  \nonumber \\
{|C^{DK}| \over |T^{DK}|} &=& 0.50 ~.
\label{DKcentral}
\end{eqnarray}
Further, the 1$\sigma$ region (whose boundary is shown as the ellipse in Fig. 1)
obtained from the $\chi^2$ analysis (See Sec. IV for more detailed discussion)
indicates that
\begin{eqnarray}
&& 0.03 \leq \cos(\delta_T -\delta_C)^{DK} \leq 0.73 ~, ~~~~
 0.42 \leq {| C^{DK}| \over |T^{DK}|} \leq 0.56 ~,
\end{eqnarray}
where the possibility that $(\delta_T -\delta_C)^{DK}=0$ is excluded.
We also note that the best fit values (shown as the black dot in Fig. 1)
with $\chi^2_{min} / d.o.f. =0.19 /1$ are
\begin{eqnarray}
(\delta_T -\delta_C)^{DK} = 71.3^\circ ~, ~~~~
 {|C^{DK}| \over |T^{DK}|} = 0.49 ~,
\end{eqnarray}
which are in good agreement with those obtained for the central values of the data
in Eq.~(\ref{DKcentral}).

\begin{figure}
    \centerline{ \DESepsf(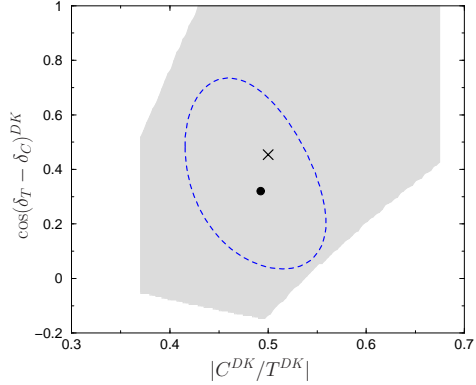 width 6cm)}
    \caption{For $B \to DK$ decays, $\cos(\delta_T -\delta_C)^{DK}$
     versus $|C^{DK} / T^{DK}|$.
     The mark ``x'' in the center denotes the result obtained from the central
     values of the experimental data.
     [The black dot is obtained from the $\chi^2$ fit with
     $\chi^2_{min}/d.o.f. =0.19 /1$ (See Sec. IV).
     The ellipse corresponds to the $(\chi^2_{min} +1)$ case.] }
\end{figure}

The strong phase difference is quite sizable.
It is also interesting to note that the contribution from the color-suppressed
tree diagram could be larger than the previously estimated one,
e.g., $|C^{DK} / T^{DK}| \approx 0.2$ given in Ref. \cite{Xing:2003fe}.
In other works, the large color-suppressed tree contribution is favored by
the present experimental data.

\begin{figure}
    \centerline{ \DESepsf(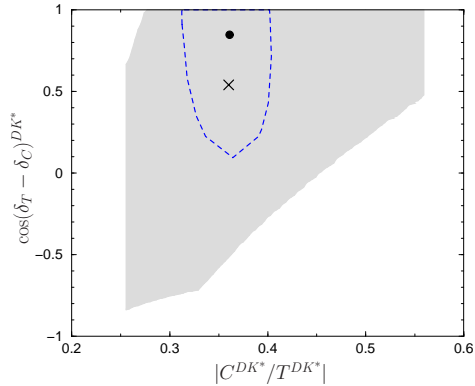 width 6cm)}
    \caption{For $B \to DK^*$ decays, $\cos(\delta_T -\delta_C)^{DK^*}$
     versus $|C^{DK^*} / T^{DK^*}|$.
     The mark ``x'' in the center denotes the result obtained from the central
     values of the experimental data.
     [The black dot is obtained from the $\chi^2$ fit with
     $\chi^2_{min}/d.o.f. =0.17 /1$ (See Sec. IV).
     The half ellipse corresponds to the $(\chi^2_{min} +1)$ case.] }
\end{figure}

For $B \to D K^*$ modes, we present $\cos (\delta_T -\delta_C)^{DK^*}$
versus $|C^{DK^*}/T^{DK^*}|$ in Fig. 2, and show the magnitudes of $T^{DK^*}$
and $C^{DK^*}$ in Table II.
For the central values of the data,
\begin{eqnarray}
(\delta_T -\delta_C)^{D K^*} &=&  57.3^\circ ~, ~~{\rm or}~~ 302.7^\circ ~, \nonumber \\
{|C^{D K^*}| \over |T^{D K^*}|} &=& 0.36 ~.
\end{eqnarray}
As in the case of $B \to DK$ decays, we obtain a similar result for $B \to DK^*$
decays: the phase difference is sizable and the large color-suppressed tree
contribution is favored.

Now let us examine the validity of the factorization approximation in
$B \to D K$ decays.
In the naive factorization approximation, the topological amplitudes
$T^{DK}$ and $C^{DK}$ are given by
\begin{eqnarray}
T^{DK} &=& {G_F \over \sqrt{2}} (V_{cb} V_{us}^*) a_1^{eff}
 \langle K^- | \bar s \gamma^{\mu} (1 -\gamma_5) u |0 \rangle
 \langle D^+ | \bar c \gamma^{\mu} (1 -\gamma_5) b |\bar B^0 \rangle
 \nonumber \\
&=& i {G_F \over \sqrt{2}} (V_{cb} V_{us}^*) a_1^{eff}
 (m_B^2 -m_D^2) f_K F_0^{B \to D} (m_K^2) ~, \nonumber \\
C^{DK} &=& {G_F \over \sqrt{2}} (V_{cb} V_{us}^*) a_2^{eff}
 \langle D^0 | \bar c \gamma^{\mu} (1 -\gamma_5) u |0 \rangle
 \langle \bar K^0 | \bar s \gamma^{\mu} (1 -\gamma_5) b |\bar B^0 \rangle
 \nonumber \\
&=& i {G_F \over \sqrt{2}} (V_{cb} V_{us}^*) a_2^{eff}
 (m_B^2 -m_K^2) f_D F_0^{B \to K} (m_D^2) ~,
\end{eqnarray}
where $V_{cb}$ and $V_{us}$ are the relevant CKM matrix elements, and
$a_1^{eff}$ and $a_2^{eff}$ are the effective Wilson coefficients.
$f_{K(D)}$ and $F_0^{B \to K(D)} (m_{D(K)}^2)$ denote the
decay constant of a $K(D)$ meson and the hadronic form factor
for the $B \to K(D)$ transition at $q^2 \equiv (p_B -p_{K(D)})^2 =m_{D(K)}^2$,
respectively.
We obtain
\begin{eqnarray}
\left| {a_2^{eff} \over a_1^{eff}} \right|
 = {1 \over r^{^{DK}}} \left| {C^{DK} \over T^{DK}} \right|
 = {1 \over r^{^{DK}}} \sqrt{{\cal B}^{DK}_{00} \over {\cal B}^{DK}_{-+}}~,
\end{eqnarray}
where
\begin{eqnarray}
r^{DK} \equiv {(m_B^2 -m_K^2) f_D F_0^{B \to K} (m_D^2)
 \over (m_B^2 -m_D^2) f_K F_0^{B \to D} (m_K^2)} ~.
\end{eqnarray}
For the central values of the data, $r^{DK}=0.81$ and $\sqrt{{\cal B}^{DK}_{00}
\over {\cal B}^{DK}_{-+}} = 0.50$, which lead to
$\left| {a_2^{eff} \over a_1^{eff}} \right| = 0.62$.
For the 1$\sigma$ range of the experimental values of the BRs, we find
\begin{equation}
0.46 \leq \left| {a_2^{eff} \over a_1^{eff}} \right| \leq 0.84 ~.
\end{equation}
For comparison, in the PQCD approach \cite{Keum:CKM2005}, it is estimated that
\begin{eqnarray}
\left| {a_2^{eff} \over a_1^{eff}} \right| \sim 0.38 ~.
\end{eqnarray}
In QCD factorization, the effective Wilson coefficient $a_1^{eff}$ for $B \to DK$
modes is the same as that for $B \to D\pi$ modes, to a good approximation \cite{Beneke:2000ry}.
The $a_1^{eff}$ for $B \to D\pi$ is presented in next section.
It is known \cite{Beneke:2000ry} that in this approach $|a_2^{eff}|$ can not be reliably
calculated, because the mechanism of color transparency is not operative for the
class-2 decays, such as $\bar B^0 \to D^0 \pi^0$ and $\bar B^0 \to D^0 \bar K^0$,
where the emission particle is a heavy charm meson.  In next section, an illustrative
value of $| {a_2^{eff} / a_1^{eff}} |$ is shown.

Similarly, for $B \to D K^*$ decays, the amplitudes $T^{DK^\ast}$ and
$C^{DK^\ast}$ can be written as
\begin{eqnarray}
&& T^{DK^\ast} = \frac{G_F}{\sqrt{2}}\left( V_{cb} V_{us}^\ast \right)
 a_1^{eff}
 \langle K^{\ast -}| \bar{s} \gamma^\mu(1- \gamma_5) u | 0 \rangle
 \langle D^+| \bar{c} \gamma_\mu(1- \gamma_5) b | \bar{B}^0 \rangle ~,
\nonumber \\
&& C^{DK^\ast} = \frac{G_F}{\sqrt{2}}\left( V_{cb} V_{us}^\ast \right)
 a_2^{eff}
 \langle D^0| \bar{c} \gamma^\mu(1- \gamma_5) u | 0 \rangle
 \langle \bar{K}^{\ast 0}| \bar{s} \gamma_\mu(1- \gamma_5) b | \bar{B}^0 \rangle ~.
\end{eqnarray}
We use the following parametrization \cite{Ball:2003rd}:
\begin{eqnarray}
&& \langle 0 | V_\mu | K^\ast \rangle = f_{K^\ast} m_{K^\ast} \epsilon_{K^\ast \mu} ~,
\nonumber \\
&&\langle \bar{K}^{\ast 0} | \bar{s} \gamma_\mu ( 1-\gamma_5) b|\bar{B}^0 \rangle
= \epsilon_{\mu\nu\rho\sigma} \epsilon_{K^\ast}^{\ast \nu} p_B^\rho
p_{K^\ast}^\sigma \frac{2 V^{B\to K^\ast} (p_D^2) }{m_B +m_{K^\ast}}
\nonumber \\
&& \hspace{4cm}
- i \epsilon_{K^\ast \mu}^\ast (m_B + m_{K^\ast})
A_1^{B \to K^\ast} (p_D^2)
\nonumber \\
&& \hspace{4cm}
+i (p_B + p_{K^\ast})_\mu \epsilon_{K^\ast}^\ast \cdot p_B
\frac{A_2^{B\to K^\ast} (p_D^2)}{m_B +m_K^\ast}
\nonumber \\
&& \hspace{4cm}
+ i p_{D\mu} \epsilon_{K^\ast}^{\ast} \cdot p_B \frac{2 m_{K^\ast}}{p_D^2}
\left( A_3^{B\to K^\ast} (p_D^2) - A_0^{B\to K^\ast} (p_D^2) \right) ~,
\end{eqnarray}
where $f_{K^\ast}$ and $\epsilon_{K^\ast}$ denote the decay constant and the
polarization vector of the $K^*$ meson, respectively.  $V^{B\to K^\ast}$ and
$A_i^{B \to K^\ast}$ ($i =0,1,2,3$) are the form factors for the $B\to K^*$
transition and given by the QCD sum rules on the light-cone  \cite{Ball:2003rd}.
With these form factors, we obtain
\begin{equation}
\left| {a_2^{eff} \over a_1^{eff}} \right|
 = {1 \over r^{^{DK^*}}} \left| {C^{DK^*} \over T^{DK^*}} \right|
 = {1 \over r^{^{DK^*}}} \sqrt{{\cal B}^{DK^*}_{00} \over {\cal B}^{DK^*}_{-+}} ~.
\end{equation}
In the $B$ rest frame, $r^{^{DK^*}}$ is given by $r^{^{DK^*}} = {a \over b}$,
where
\begin{eqnarray}
&&
a = \left| \left\{ -(m_B+m_{K^\ast}) A_1^{B\to K^\ast} (m_D^2)
 + (m_B - m_{K^\ast}) A_2^{B\to K^\ast} (m_D^2) \right. \right.
\nonumber \\
&& \hspace{1cm}
 \left. \left.
 + 2 m_{K^\ast} \left( A_3^{B \to K^\ast } (m_D^2) - A_0^{B\to K^\ast} (m_D^2)
 \right) \right\} ( p_B \cdot \epsilon_{K^\ast}^\ast ) \right| ~,
\nonumber \\
&&
b = \frac{\sqrt{m_B m_D} (m_B +m_D)}{2 m_B m_D}
 f_{K^\ast} \xi_D \lambda^{1/2} (m_B^2, m_D^2, m_{K^\ast}^2) ~.
\end{eqnarray}
Here $\xi_D$ is the Isgur-Wise function and
$\lambda(m_B^2, m_D^2, m_{K^\ast}^2)
= m_B^4 + m_D^4 + m_{K^\ast}^4 - 2m_B^2 m_D^2 -2m_B^2 m_{K^\ast}^2
-2 m_D^2 m_{K^\ast}^2$.
Using the central values of the data, we obtain $r^{^{DK^*}}=0.82$ and
$\sqrt{{\cal B}^{DK^*}_{00} \over {\cal B}^{DK^*}_{-+}}=0.36$, which give
$\left| {a_2^{eff} \over a_1^{eff}} \right| = 0.44$.
For the $1\sigma$ range of the experimental data, the allowed value of
the ratio $| a_2^{eff} / a_1^{eff} |$ is in between 0.31 and 0.69.

\mbox{}From the above results, we see that if one assumes the
naive factorization in $B \to DK$ and $B \to D K^*$ decays, the
favored value of the ratio $| a_2^{eff} / a_1^{eff} |$ is much
larger than the usual estimate $| a_2^{eff} / a_1^{eff} | \sim
0.25$ \cite{Xing:2003fe}. This can be possibly understood if the
magnitude of the color-suppressed tree amplitude $C^{DK^{(*)}}$ is
effectively enhanced due to non-factorizable contributions as in
the PQCD approach \cite{Keum:2003js} or final-state interactions
\cite{Cheng:2004ru}.

\section{$B \to D \pi$ and $D \rho$ decay modes}

Let us turn to $B \to D \pi$ decays.
The decay amplitudes can be represented in terms of the topological
amplitudes $T^{D\pi}$, $C^{D\pi}$ and $E^{D\pi}$:
\begin{eqnarray}
A^{D\pi}_{0-} &\equiv& {\cal A}(B^- \to D^0 \pi^-)
 = T^{D\pi} + C^{D\pi} ,  \nonumber \\
A^{D\pi}_{+-} &\equiv& {\cal A}(\bar B^0 \to D^+ \pi^-)
 = T^{D\pi} + E^{D\pi} ,  \nonumber \\
\sqrt{2} A^{D\pi}_{00} &\equiv& \sqrt{2} {\cal A}(\bar B^0 \to D^0 \pi^0)
 = -C^{D\pi} + E^{D\pi} ,
\label{DPIdecayandtopo}
\end{eqnarray}
where the topological amplitudes $T^{D\pi}$, $C^{D\pi}$ and $E^{D\pi}$
are defined as
\begin{eqnarray}
X^{D\pi} \equiv |X^{D\pi}| e^{i \delta'_X}
 \equiv |V_{cb} V_{ud}^*| a'_X e^{i \delta_X},  ~~ (X = T,C,E)
\label{DPItopoamp}
\end{eqnarray}
with the real amplitude $a'_{T(C,E)}$ and the strong phases $\delta'_{T(C,E)}$.
The above amplitudes involve no weak phase because of the CKM factor
$V_{cb} V_{ud}^*$.

In the above equations (\ref{DPIdecayandtopo}) and (\ref{DPItopoamp}),
there are five unknown parameters
(one relative phase can be removed), while only three BRs have been measured from
experiments.  To determine the unknown parameters, one needs more
information.  For this purpose, one may invoke flavor SU(3) symmetry
to connect the amplitudes for $B \to D\pi$ to those for $B \to DK$.
However, as we shall see later, the SU(3) breaking effect can be sizable.
Therefore, in our analysis, instead of using the SU(3) symmetry between
all the relevant amplitudes, we use the SU(3) symmetry only for the
exchange amplitude $E^{D\pi}$, because the exchange contribution
is expected to be small due to a suppression factor of $f_B / m_B$.
\mbox{}From the measured BR for $\bar B^0 \to D_s^+ K^-$ which involves only
the $W$-exchange diagram, we obtain $|E^{D_s K}| =(0.71 \pm 0.10)
\times 10^{-7}$ GeV \cite{Chiang:2002tv}.
To be even more conservative, considering the SU(3)
breaking effect, we allow that $|E^{D\pi}|$ lies within the 2$\sigma$
range, which leads to $|E^{D\pi}| =(0.71 \pm 0.20) \times 10^{-7}$ GeV.
Further, we allow that $(\delta'_T - \delta'_E)^{D\pi}$ can vary
from 0 to $2\pi$.

The amplitudes and the phase differences can be written as
\begin{eqnarray}
|T^{D\pi}| &=& \sqrt{{8 \pi m_B^2 \over p_{_{D\pi}} \tau_0} {\cal B}^{D\pi}_{+-}
 - |E^{D\pi}|^2 \sin^2 (\delta'_T -\delta'_E)^{D\pi}}
 - |E^{D\pi}| \cos(\delta'_T -\delta'_E)^{D\pi},
 \nonumber \\
|C^{D\pi}| &=& \sqrt{{16 \pi m_B^2 \over p_{_{D\pi}} \tau_0} {\cal B}^{D\pi}_{00}
 - |E^{D\pi}|^2 \sin^2 (\delta'_C -\delta'_E)^{D\pi}}
 + |E^{D\pi}| \cos(\delta'_C -\delta'_E)^{D\pi},
\nonumber \\
\cos(\delta'_T -\delta'_C)^{D\pi} &=& {1 \over 2 |T^{D\pi}| |C^{D\pi}|}
 \left[ {{8 \pi m_B^2 \over p_{_{D\pi}} \tau_-} {\cal B}^{D\pi}_{0-}
 - |T^{D\pi}|^2 - |C^{D\pi}|^2} \right],
\nonumber \\
\cos(\delta'_C -\delta'_E)^{D\pi} &=& -{1 \over 2 |C^{D\pi}| |E^{D\pi}|}
 \left[ {{16 \pi m_B^2 \over p_{_{D\pi}} \tau_0} {\cal B}^{D\pi}_{00}
 - |C^{D\pi}|^2 - |E^{D\pi}|^2} \right] .
\end{eqnarray}
The above relations hold for the corresponding $B \to D\rho$ decay modes as well
and can be used for the relevant $B \to D\rho$ decays by simply replacing
$\pi$ by $\rho$.

\begin{figure}
   \centerline{ \DESepsf(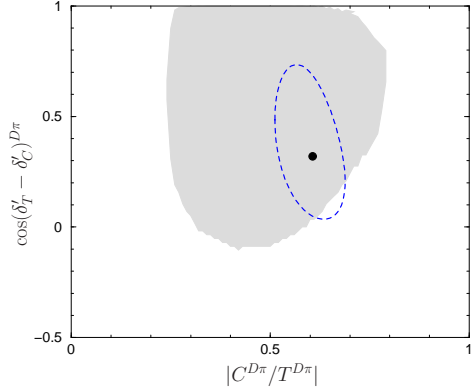 width 6cm)}
    \caption{For $B \to D\pi$ decays, $\cos(\delta'_T -\delta'_C)^{D\pi}$
     versus $|C^{D\pi} / T^{D\pi}|$.
     [The black dot is obtained from the $\chi^2$ fit with
     $\chi^2_{min}/d.o.f. =0.19 /1$ (See Sec. IV).
     The ellipse corresponds to the $(\chi^2_{min} +1)$ case.] }
\end{figure}

Similarly to the case of $B \to DK$ decays, using the experimental result
for $B \to D\pi$, we compute the magnitudes of the relevant amplitudes and
the phase differences.
Our numerical result is shown in Fig. 3 as a graph of
$\cos(\delta'_T -\delta'_C)^{D\pi}$ versus $|C^{D\pi} / T^{D\pi}|$.
The magnitudes of $T^{D\pi}$ and $C^{D\pi}$ are shown in Table II.
The best fit values (shown as the black dot in Fig. 3) with
$\chi^2_{min} / d.o.f. =0.19 /1$ are (See Sec. IV)
\begin{eqnarray}
&& |T^{D\pi}| = 5.85 \times 10^{-7} {\rm GeV},
 ~~~ |C^{D\pi}| = 3.56 \times 10^{-7} {\rm GeV},
\nonumber \\
&& |E^{D\pi}| = 0.86 \times 10^{-7} {\rm GeV},
 ~~~ (\delta'_T -\delta'_C)^{D\pi} = 71.3^{\circ} ~.
\label{DPIchi2}
\end{eqnarray}

\begin{figure}
   \centerline{ \DESepsf(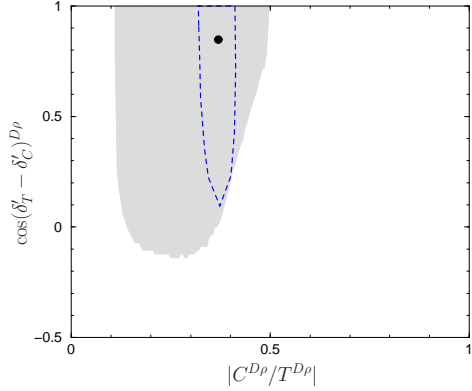 width 6cm)}
    \caption{For $B \to D\rho$ decays, $\cos(\delta'_T -\delta'_C)^{D\rho}$
     versus $|C^{D\rho} / T^{D\rho}|$.
     [The black dot is obtained from the $\chi^2$ fit with
     $\chi^2_{min}/d.o.f. =0.17 /1$ (See Sec. IV).
     The half ellipse corresponds to the $(\chi^2_{min} +1)$ case.] }
\end{figure}

For $B \to D\rho$ decays, we also obtain similar results.
In this case, since only the upper bound for the BR of $\bar B^0 \to D_s^+ K^{*-}$
(which involves only the annihilation contribution)
is known at present, we use $|E^{D\rho}| =(0.71 \pm 0.20) \times 10^{-7}$ GeV
as in the case of $B \to D\pi$.  As we shall see in Sec. IV this treatment turns
out to be reasonable.
We present the graph of $\cos(\delta'_T -\delta'_C)^{D\rho}$ versus
$|C^{D\rho} / T^{D\rho}|$ in Fig. 4.
The best fit values (shown as the black dot in Fig. 4) with
$\chi^2_{min} / d.o.f. = 0.17 /1$ are (See Sec. IV)
\begin{eqnarray}
&& |T^{D\rho}| = 9.57 \times 10^{-7} {\rm GeV},
 ~~~ |C^{D\rho}| = 3.55 \times 10^{-7} {\rm GeV},
\nonumber \\
&& |E^{D\rho}| = 0.75 \times 10^{-7} {\rm GeV},
 ~~~ (\delta_T -\delta_C) = 32.4^{\circ} ~.
\end{eqnarray}

Let us turn to examine the validity of the factorization in $B \to D\pi$
and $B \to D\rho$ decays.
For $B \to D\pi$ decays, neglecting the small $E^{D\pi}$,
\begin{eqnarray}
\left| {a_2^{eff} \over a_1^{eff}} \right|
 = {1 \over r^{^{D\pi}}} \left| {C^{D\pi} \over T^{D\pi}} \right|
 = {1 \over r^{^{D\pi}}} \sqrt{2 {\cal B}^{D\pi}_{00} \over {\cal B}^{D\pi}_{-+}}
 = 0.54 \sim 0.70 ~,
\end{eqnarray}
where
\begin{eqnarray}
r^{^{D\pi}} \equiv {1 \over \sqrt{2}}
 {(m_B^2 -m_{\pi}^2) f_D F_0^{B \to \pi} (m_D^2)
 \over (m_B^2 -m_D^2) f_{\pi} F_0^{B \to D} (m_{\pi}^2)}
 = 0.54 ~.
\end{eqnarray}
For comparison, in PQCD calculation \cite{Keum:CKM2005}, it is predicted that
\begin{eqnarray}
\left| {a_2^{eff} \over a_1^{eff}} \right| = 0.42 \sim 0.50 ~,
\end{eqnarray}
when the contribution from the exchange diagrams is neglected, and
\begin{eqnarray}
\left| {a_2^{eff} \over a_1^{eff}} \right| = 0.37 \sim 0.45 ~,
\end{eqnarray}
when the contribution from the exchange diagrams is included.
In the QCD factorization approach \cite{Beneke:2000ry}, $|a_1^{eff}|$ is estimated
as $|a_1^{eff}| \approx 1.05$.  But, as commented in the previous section,
in this approach $|a_2^{eff}|$ can not be reliably estimated.
For an illustration, a rough estimation \cite{Beneke:2000ry} shows
\begin{eqnarray}
\left| {a_2^{eff} \over a_1^{eff}} \right| \sim 0.24 ~.
\end{eqnarray}

For $B \to D\rho$ decays, neglecting the small $E^{D\rho}$,
\begin{eqnarray}
\left| {a_2^{eff} \over a_1^{eff}} \right|
 = {1 \over r^{^{D\rho}}} \left| {C^{D\rho} \over T^{D\rho}} \right|
 = {1 \over r^{^{D\rho}}} \sqrt{2 {\cal B}^{D\rho}_{00} \over {\cal B}^{D\rho}_{-+}}
 = 0.24 \sim 0.42 ~.
\end{eqnarray}
In the $B$ rest frame, $r^{D\rho}$ is given by
$r^{D\rho} = {a' \over b'} =0.60$,
where
\begin{eqnarray}
&&
a' =\frac{1}{\sqrt{2}} \left| \left\{ -(m_B+m_{\rho}) A_1^{B\to \rho} (m_D^2)
+ (m_B - m_{\rho}) A_2^{B\to \rho} (m_D^2)
\right. \right.
\nonumber \\
&& \hspace{1cm}
\left. \left.
+ 2 m_{\rho} \left( A_3^{B \to \rho } (m_D^2) - A_0^{B\to \rho} (m_D^2)
\right) \right\} (p_B \cdot \epsilon_{\rho}^\ast) \right| ~,
\nonumber \\
&&
b' = \frac{\sqrt{m_B m_D} (m_B +m_D)}{2 m_B m_D}
f_{\rho} \xi_D \lambda^{1/2} (m_B^2, m_D^2, m_{\rho}^2) ~.
\end{eqnarray}
As in the cases of $B \to DK$ and $B \to DK^*$, the large values of
$| a_2^{eff} / a_1^{eff} |$ are favored for $B \to D\pi$ and $B \to D\rho$
decays.  It indicates that for $B \to D\pi$ and $B \to D\rho$ the
color-suppressed tree contributions to $C^{D\pi}$ and $C^{D\rho}$ are
effectively enhanced.  The possible mechanism for this enhancement is
either the short-distance non-factorizable contribution \cite{Keum:2003js}
or large final-state rescattering interactions \cite{Cheng:2004ru},
or both of them.

\section{The $\chi^2$ analysis using $B \to DK$, $D\pi$, and $B \to D K^*$, $D\rho$}

In order to find the most likely values of the magnitudes of the topological
amplitudes and the strong phase shifts in $B \to DK$, $D\pi$, and $DK^*$, $D\rho$,
we do the $\chi^2$ analysis using the BRs of these decay processes.
First we assume the flavor SU(3) symmetry between the topological amplitudes
for $B \to DK$ and $D \pi$ (similarly for $B \to D K^*$ and $D\rho$).
Then we will take into account the SU(3) breaking effect.

\subsection{The $B \to DK$ and $D\pi$ case}

Assuming the flavor SU(3) symmetry,
we have six observables (the measured BRs of $B \to DK$ and $D\pi$, as shown
in Table I) and five parameters [$|T^{DK}|$, $|C^{DK}|$, $|E^{D\pi}|$, $(\delta_T -\delta_C)$,
$(\delta_T -\delta_E)$] so that the degrees of freedom ($d.o.f.$) for the fit
is 1.   Without considering the SU(3) breaking effect,
$T^{D\pi}$ and $C^{D\pi}$ are given by
$T^{D\pi} =T^{DK} \left( {V_{ud} \over V_{us}} \right)$ and
$C^{D\pi} =C^{DK} \left( {V_{ud} \over V_{us}} \right)$, respectively.
In this case we find that $\chi^2_{min} / d.o.f =3.34 /1$ indicating a poor fit.
Taking into account the SU(3) breaking at first order, such as
$T^{D\pi} = T^{DK} \left( {V_{ud} f_{\pi} \over V_{us} f_K} \right)$
and $C^{D\pi} = C^{DK} \left( {V_{ud} \over V_{us}} \right)$, we find the best
fit with $\chi^2_{min} / d.o.f. =0.19 /1$.
The corresponding parameter values are
\begin{eqnarray}
&& |T^{DK}| = 1.64 \times 10^{-7} {\rm GeV}, ~~~ |C^{DK}| = 0.81 \times 10^{-7} {\rm GeV},
~~~ |E^{D\pi}| = 0.86 \times 10^{-7} {\rm GeV},   \nonumber \\
&& (\delta_T -\delta_C) = 71.3^{\circ} ~, ~~~~~ (\delta_T -\delta_E) = 91.2^{\circ} ~.
\end{eqnarray}
The $\cos (\delta_T -\delta_C)$ versus $|C/T|$ obtained from the above
ones for $\chi^2_{min}$ is depicted for $B \to DK$ in Fig. 2 and for $B \to D\pi$
in Fig. 4.  The result for $(\chi^2_{min} +1)$ is also shown as an ellipse
in the same figures and their numerical values are shown in Table II.

For the best fit, we find that $\left| {C^{DK} \over T^{DK}} \right| = 0.49$,
which indicates the relatively large color-suppressed tree contribution.
The best fit value for $|E^{D\pi}|$ is in good agreement with
$|E^{D\pi}| =(0.71 \pm 0.20) \times 10^{-7}$ GeV used in Sec. III.
Our result indicates that the exchange contribution in $B \to D \pi$ decay
can be sizably enhanced as well, which is contrary to the usual estimate
in the QCD factorization.
Notice that within the flavor SU(3) symmetry with a reasonable SU(3) breaking
effect in the $B\to DK$ and $D\pi$ decays,
the strong phase difference between the color-{\it allowed} and -{\it suppressed}
decay amplitudes does not vanish at the level of one standard deviation.

\subsection{The $B \to D K^*$ and $D\rho$ case}

Similarly to the $B \to DK$ and $D\pi$ case, we have the six measured BRs of
$B \to D K^*$ and $D\rho$ and the five parameters.
Assuming the flavor SU(3) symmetry [$T^{D\rho} = T^{DK^*}
\left( {V_{ud} \over V_{us}} \right)$], we find that
$\chi^2_{min} / d.o.f. =0.10 /1$.
The corresponding parameters are
\begin{eqnarray}
&& |T^{DK^\ast}| = 2.24 \times 10^{-7} {\rm GeV},
~~~ |C^{DK^\ast}| = 0.81 \times 10^{-7} {\rm GeV},
~~~ |E^{D\rho}| = 0.86 \times 10^{-7} {\rm GeV},   \nonumber \\
&& (\delta_T -\delta_C) = 39.7^{\circ} ~,
~~~~~ (\delta_T -\delta_E) = 65.3^{\circ} ~.
\end{eqnarray}
Taking into account the SU(3) breaking effect [$T^{D\rho} = T^{DK^*}
\left( {V_{ud} f_{\rho} \over V_{us} f_{K^*}} \right)$ and
$C^{D\rho} = C^{DK^*} \left( {V_{ud} \over V_{us}} \right)$], we find
another good fit with $\chi^2_{min} / d.o.f. = 0.17 /1$.  The corresponding
parameters in this case are
\begin{eqnarray}
&& |T^{DK^\ast}| = 2.23 \times 10^{-7} {\rm GeV},
~~~ |C^{DK^\ast}| = 0.81 \times 10^{-7} {\rm GeV},
~~~ |E^{D\rho}| = 0.75 \times 10^{-7} {\rm GeV},   \nonumber \\
&& (\delta_T -\delta_C) = 32.4^{\circ} ~,
~~~~~ (\delta_T -\delta_E) = 34.8^{\circ} ~.
\end{eqnarray}
The numerical values of $|C/T|$ and $\cos(\delta_T -\delta_C)$ for
$(\chi^2_{min} +1)$ are shown in Table II.

Unlike the $B \to DK$ and $D\pi$ case, good fits are obtained for both cases of
the SU(3) symmetry and the broken SU(3) symmetry.
This can be understood that in the $B \to DK^\ast$ and $D \rho$ case the flavor
SU(3) breaking factor $f_\rho/f_{K^\ast}$ is almost equal to unity, while
in the  $B \to DK$ and $D\pi$ case the breaking factor $f_\pi/f_K$ is relatively
large.
Further, the parameters obtained in the case of the SU(3) symmetry are quite
similar to those obtained in the broken SU(3) symmetry, except the parameter
$(\delta_T -\delta_E)$ which shows a sizable difference in the two cases.
The black dots in Figs. 3 (for $B \to DK^*$) and 4 (for $B \to D\rho$) show
$\cos (\delta_T -\delta_C)$ versus $|C/T|$ for $\chi^2_{min} / d.o.f. = 0.17 /1$.
Those values for $(\chi^2_{min} +1)$ are shown as a half ellipse in the
same figures.

We notice that the relatively large color-suppressed tree contribution is favored
in the $B \to DK^*$ and $D\rho$ case as well: $\left| {C \over T} \right| = 0.36$
for $\chi^2_{min} / d.o.f. = 0.17 /1$.  The magnitude of $E^{D\rho}$ is also
consistent with the one used in Sec. III.
It implies that the BR for $\bar B^0 \to D_s^+ K^{*-}$ would be similar to that
for $\bar B^0 \to D_s^+ K^-$.  This will be tested with future experimental results
on the BRs for these decay modes.

\section{Flavor SU(3) symmetry breaking effect}

Let us estimate the flavor SU(3)symmetry breaking effect in $B \to DK ~(DK^*)$
and $B \to D\pi ~(D\rho)$ decays.
If flavor SU(3) were exact, one would get for $B \to DK$ and $D\pi$,
\begin{equation}
\left|{ T^{DK} \over V_{cb} V^*_{us} } \right|
 = \left|{ T^{D\pi} \over V_{cb} V^*_{ud} } \right| ~, ~~~
\left|{ C^{DK} \over V_{cb} V^*_{us} } \right|
 = \left|{ C^{D\pi} \over V_{cb} V^*_{ud} } \right| ~,
\end{equation}
and for $B \to DK^*$ and $D\rho$,
\begin{equation}
\left|{ T^{DK^*} \over V_{cb} V^*_{us} } \right|
 = \left|{ T^{D\rho} \over V_{cb} V^*_{ud} } \right| ~, ~~~
\left|{ C^{DK^*} \over V_{cb} V^*_{us} } \right|
 = \left|{ C^{D\rho} \over V_{cb} V^*_{ud} } \right| ~.
\end{equation}

To estimate the SU(3) breaking effect, let us take the central values
of the data as a typical example.
We find for $B \to DK$ and $B \to D\pi$,
\begin{equation}
\left| {{T^{DK} /(V_{cb} V^*_{us})} \over
 {T^{D\pi} /(V_{cb} V^*_{ud})}} \right| = 1.21 ~, ~~
\left| {{C^{DK} /(V_{cb} V^*_{us})} \over
 {C^{D\pi} /(V_{cb} V^*_{ud})}} \right| = 1.29 ~,
\label{SU3breakTDKDpi}
\end{equation}
and for $B \to DK^*$ and $B \to D\rho$,
\begin{equation}
\left| {{T^{DK^*} /(V_{cb} V^*_{us})} \over
 {T^{D\rho} /(V_{cb} V^*_{ud})}} \right| = 0.96 ~, ~~
\left| {{C^{DK^*} /(V_{cb} V^*_{us})} \over
 {C^{D\rho} /(V_{cb} V^*_{ud})}} \right| = 1.27 ~.
\label{SU3breakTDKstarDrho}
\end{equation}
The above result shows that the SU(3) breaking effect can be sizable: i.e.,
about $(20 - 30)\%$ at the amplitude level, except the color-allowed tree
amplitudes for $B \to DK^*$ and $D\rho$.
Our result for the color-allowed tree amplitudes in (\ref{SU3breakTDKDpi})
and (\ref{SU3breakTDKstarDrho}) agrees with that of Ref. \cite{Xing:2001nj}.
But, the result for the color-suppressed tree amplitudes
does not agree with the estimate in the naive factorization shown in
\cite{Xing:2001nj} and shows about two or three times larger breaking effect.
It again indicates that the color-suppressed tree amplitudes can not
be reasonably estimated by the naive factorization, because they can be
effectively enhanced by non-factorizable effect and final-state interactions,
as discussed before.

\section{Conclusion}

We studied $B \to DK$, $DK^*$ and $B \to D\pi$, $D\rho$ decay processes in
a model-independent way.  Using the quark diagram decomposition
of the decay amplitudes and the present experimental result on the relevant
BRs, we determined the magnitudes and the relative strong phase shifts of
the relevant amplitudes.

First we analyzed the $B \to DK^{(*)}$ and $B \to D\pi ~(D\rho)$
modes separately from each other so that the flavor SU(3) symmetry is not needed
to combine the relevant amplitudes in $B \to DK^{(*)}$ and $B \to D\pi ~(D\rho)$
with each other.  As shown in Sec. V, the SU(3) breaking effect can be
sizable in these modes.
Further, in order to determine the most likely values for the relative strong
phases and the magnitudes of the amplitudes in a statistically reliable way, we used
the $\chi^2$ minimization technique.  In this case, we used the flavor SU(3)
symmetry, but took its breaking effect into account as well.

Our results show that the strong phase differences between the
color-allowed and the color-suppressed tree amplitudes can be
large: for instance, for the $B \to DK ~ (D\pi)$ mode, the best
fit value for $(\delta_T -\delta_C)$ is $71.3^{\circ}$. It should
be emphasized that $(\delta_T -\delta_C)$ is {\it non-zero} at
1$\sigma$ level (Figs. 1 and 3).  This result is obtained from the
statistical approach and clearly different from those of the
previous works, where $(\delta_T -\delta_C)$ was assumed to be
$0^\circ$ \cite{Xing:2001nj,Xing:2003fe}, or the vanishing
$(\delta_T -\delta_C)$ could not be excluded with the present data
\cite{Chiang:2002tv}.

Another interesting result is that in $B \to DK^{(*)}$, $D\pi$,
$D\rho$ decays, the color-suppressed tree contributions are
effectively enhanced, which is inconsistent with the naive
expectation in the factorization approximation. For example, the
best fit value is $|C / T| = 0.49$ for $B \to DK$, and $|C / T| =
0.60$ for $B \to D\pi$. These ratios are quite larger than
previously estimated ones as in \cite{Xing:2001nj,Xing:2003fe},
but are consistent with the recent results as in
\cite{Keum:2003js,Cheng:2004ru,Mantry:2003uz}.

\vspace{1cm}
\centerline{\bf ACKNOWLEDGEMENTS}
\noindent
The work of C.S.K. was supported
in part by  CHEP-SRC Program,
in part by Grant No. R02-2003-000-10050-0 from BRP of the KOSEF
and in part by 2004 Yonsei University Grant.
The work of S.O. was supported by Korea Research Foundation Grant
(KRF-2004-050-C00005).
The work of C.Y. was supported in part by Brain Korea 21 Program
and in part by Grant No. F01-2004-000-10292-0 of KOSEF-NSFC International
Collaborative Research Grant.


\end{document}